\documentclass[11pt]{article}
\newlength{\defaultparindent}
\def\R{{\mathbb{R}}} % good; MB XII 2005
\def\C{{\mathbb{C}}} % good; MB XII 2005
\def\Identity{{\mathbb{1}}} % package bbold; to be tried; MB XI 2005

\newcommand{\JJ}{\mathbin{\raisebox{0.25ex}{$\scriptstyle % originally was \footnotesize but is "invalid in math mode" - mbh
                       \rm\vphantom{I}%
                       \_\hskip -0.25em\_%
                       \vrule width 0.6pt$}}}           %left contraction (from Makro98.tex)

\usepackage{latexsym}
\usepackage{amsfonts}
\usepackage{bbold}
\newtheorem{MS_theorem}{Theorem}
\newtheorem{MS_Proposition}{Proposition}
\def\myconjugate#1{\bar{#1}} % this way conjugate definitions can be easily changed

\begin{document}

\title{\bf A Spinorial Formulation of the Maximum Clique Problem of a Graph}

\author{\\
	\bf Marco Budinich\\
	Dipartimento di Fisica\\
	Universit\`a di Trieste \& INFN\\
	Via Valerio 2, I - 34127 Trieste, Italy\\
	\texttt{mbh@ts.infn.it}\\
	\\
	\bf Paolo Budinich\\
	International School for Advanced Studies -- SISSA/ISAS\\
	Via Beirut 4, I - 34014 Trieste, Italy\\
	\texttt{fit@ictp.trieste.it}\\
	\\
	Submitted to: Journal of Mathematical Physics
	}

\date{ December 23, 2005 }

\maketitle

\begin{abstract}
We present a new formulation of the maximum clique problem of a graph in complex space. We start observing that the adjacency matrix $A$ of a graph can always be written in the form $A = B^2$ where $B$ is a complex, symmetric matrix formed by vectors of zero length (null vectors) and the maximum clique problem can be transformed in a geometrical problem for these vectors. This problem, in turn, is translated in spinorial language and we show that each graph uniquely identifies a set of pure spinors, that is vectors of the endomorphism space of Clifford algebras, and the maximum clique problem is formalized in this setting so that, this much studied problem, may take advantage from recent progresses of pure spinor geometry.
\end{abstract}

\thispagestyle{empty}
\newpage

\section{Introduction}
In this paper we propose a new representation of the maximum clique problem in complex space. After a brief review of this famous NP-complete problem, we show how the adjacency matrix of a graph can be expressed as the square of a symmetric complex matrix. The vectors forming this matrix have zero length and Cartan has shown that this geometry can be treated elegantly with spinors. After a brief remind of spinor properties, we show that the adjacency matrix is better decomposed in the Witt basis of complex space. We finish with a formulation of the maximum clique problem in this formalism and show that each graph uniquely identifies a spinor whose properties surely deserve deeper studies.

\section{The Maximum Clique Problem}

\subsection{A brief review}

Given a graph of size $n$, a clique is a subgraph with pairwise adjacent vertices and the Maximum Clique (MC) problem is that of finding the {\em size\/} $k$ of the largest clique. It is a well studied NP-complete problem and there are reviews with hundreds references (see e.g. \cite{Pardalos 1994} or \cite{Bomze 1999}).

Given a graph let $A$ be its $n \times n$ adjacency matrix with elements in $\{0,1\}$ and zero diagonal; we will consider only undirected graphs for which $A$ is symmetric. Furthermore, since every undirected graph can be subdivided in connected graphs, we discuss only connected graphs that have irreducible adjacency matrices.

The quadratic form on $A$ ($^{\prime}$ indicates transposition, bold characters, vectors) is bounded by
\begin{displaymath}
0 \leq {\bf x}^{\prime}A{\bf x} \leq 1 - \frac{1}{n} \quad \textrm{for} \quad {\bf x} \in K_{n}
\end{displaymath}
where the simplex $K_{n} = \{{\bf x} \in \R^{n} : x_{i} \geq 0 \: \forall i$ and ${\bf e}^{\prime} {\bf x} = 1\}$ and ${\bf e}^{\prime} = (1,1,\ldots,1)$.

A subgraph with $r$ vertices is uniquely determined by its characteristic vector that is an $n$ dimensional vector whose $i$-$th$ component, by taking values $1/r$ or $0$, indicates whether the $i$-$th$ element belongs or not to the subgraph. Characteristic vectors belong to $K_{n}$.

In 1965 Motzkin and Straus \cite{Motzkin 1965} proved the following

\begin{MS_theorem}
If the MC of graph $A$ has size $k$ then
\begin{equation}
\max_{{\bf x}\in K_{n}} {\bf x}^{\prime} A{\bf x} = 1 - \frac{1}{k}
\label{formula_b}
\end{equation}
and if ${\bf x}_{k}$ is the characteristic vector of a MC then ${\bf x}_{k}^{\prime} A{\bf x}_{k} =1 - \frac{1}{k}$.
\end{MS_theorem}

Bomze \cite{Bomze 1997} sharpened this result showing that, if $\Identity$ represents the identity matrix, $\max_{{\bf x}\in K_{n}} {\bf x}^{\prime} (A + \frac{1}{2} \Identity){\bf x} = 1 - 1/2k$ and moreover that this quadratic form reaches its maximum {\em if\/} and {\em only if\/} {\bf x} is the characteristic vector of a MC.

With this formulation the, essentially combinatorial, MC problem is transposed to the search of the maximum of a quadratic function in a bounded region: a continuous optimization problem with linear constraints. Several authors \cite{Pardalos 1990, Pelillo 1995, Gibbons 1997} used this formulation to find approximate solutions to the MC problem.

\subsection{Decomposition of the adjacency matrix}

Any symmetric matrix like $A$ may be expressed in the form (see Appendix for details)
\begin{equation}
A = B^{\prime} B = B B = B^2
\label{formula_f}
\end{equation}
where $B$ is a complex, symmetric matrix that we can think as formed by $n$ complex column vectors ${\bf z}_i \in \C^{n}$:
\begin{equation}
B=({\bf z}_{1},{\bf z}_{2},\ldots,{\bf z}_{n})
\hspace{2cm}
a_{jk} = (B^{\prime} B)_{jk} = {\bf z}_{j}^{\prime} {\bf z}_{k}
\label{formula_g}
\end{equation}
these vectors are called null vectors since they have zero length since $a_{jj} ={\bf z}_{j}^{2} =0$. Let $n_0$ be the number of zero eigenvalues of $A$, then ${\rm rank} A = {\rm rank} B = n - n_0$ and this is also the dimension of the space $V$ spanned by the vectors ${\bf z}$. With $B$ the quadratic form on $A$ becomes:
\begin{displaymath}
{\bf x}^{\prime}A{\bf x} = {\bf x}^{\prime}B^{\prime}B{\bf x} = (B{\bf x})^{2} := {\bf Z}^2
\end{displaymath}
where $ {\bf Z} = B{\bf x} = \sum\nolimits_{i} x_{i} {\bf z}_{i} \in \C^n $ is a complex vector with which the MC problem may be reformulated and (\ref{formula_b}) can be written
\begin{displaymath}
\max_{{\bf x}\in K_{n}} {\bf Z}^{2} = 1 - \frac{1}{k}
\end{displaymath}
this is the problem of finding the vector of maximum length that can be formed by a convex combination of the ${\bf z}$'s%
\footnote{In real space this is the problem of finding the point(s) of the convex hull of the ${\bf z}$'s with maximum distance from the origin.}%
.
We remark that ${\bf Z}^{2}$ is real since $x_{i}$ and ${\bf z}_{j}^{\prime} {\bf z}_{k} = a_{jk} $ are all real.

\subsection{Another formulation of the maximum clique problem}

The MC problem is tightly connected to the problem of the maximum independent set that is the problem of finding the size of the largest subgraph whose vertices are pairwise nonadjacent. Let $\myconjugate{A}$ represent the adjacency matrix of the complementary graph of $A$, i.e. the graph with the same vertices and complementary edges; if $J$ is the matrix whose elements are all $1$ then $\myconjugate{A} = J - \Identity - A$. It is simple to verify that every subset of vertices of $A$ that forms a clique forms also an independent set of $\myconjugate{A}$ and vice versa. The cliques of $A$ are thus in one to one correspondence to the independent sets of $\myconjugate{A}$ and if ${\bf x}_k$ is the characteristic vector of a clique of $A$ then it is also the characteristic vector of an independent set of $\myconjugate{A}$ and ${\bf x}_{k}^{\prime} \myconjugate{A} {\bf x}_{k} = 0$.

There are several ways to formulate the MC problem of $A$ as the problem of the maximum independent set of $\myconjugate{A}$ \cite{Pardalos 1994}, \cite{Knuth 1994}. Indicating $\myconjugate{\bf Z} =\myconjugate{B} {\bf x}$ (from now on we use overstriked symbols to indicate quantities relative to $\myconjugate{A}$) a formulation with appealing properties is
\begin{equation}
\max_{{\bf x} \in \{ \{0,1\}^{n} : \myconjugate{\bf Z}^{2} = 0\}}
{\bf x}^{\prime} {\bf x} = k \; {\rm .}
\label{formula_j}
\end{equation}
This problem has the following geometrical interpretation: the null vectors ${\myconjugate{\bf z}}$ span the space $V \subseteq \C^n$ and any couple of linearly independent vectors $\myconjugate{\bf z}_{j} $ and $\myconjugate{\bf z}_{k} $ span a two-dimensional space contained in $V$. If $\myconjugate{\bf z}_{j}^{\prime} \myconjugate{\bf z}_{k} = \myconjugate{a}_{jk} = 0$ it is easy to verify that this space has the property that all of its elements are null vectors and are all mutually orthogonal: this space is called a Totally Null Plane (TNP). If $\myconjugate{A}$ contains at least one nondiagonal zero element, then $V$ contains at least one two dimensional TNP.

The solution of the MC problem provides the largest subset ${\myconjugate{\bf z}_{j_{1}} ,\myconjugate{\bf z}_{j_{2}} ,\ldots, \myconjugate{\bf z}_{j_{k}}}$ of ${\myconjugate{\bf z}}$'s which define a TNP contained in $V$. We note that, since for any two of them $\myconjugate{\bf z}_{j_{r}}^{\prime} \myconjugate{\bf z}_{j_{s}} =\myconjugate{a}_{j_{r}j_{s}} = 0$, then not only $\sum\nolimits_{l=1,k} \myconjugate{\bf z}_{j_{l}}$, but any of their linear combinations
\begin{displaymath}
\myconjugate{\bf Z} = \sum\nolimits_{l=1,k} x_{j_{l}} \myconjugate{\bf z}_{j_{l}}
\end{displaymath}
is a particular null vector that satisfy the constraint $\myconjugate{\bf Z}^{2} = 0$. Since $\C^n$ cannot contain a TNP of more than $\frac{n}{2}$ dimensions \cite{Cartan 1937} this formulation have already been used to calculate an upper bound for the size of the MC \cite{Budinich 2003}.

\section{A brief review of spinors}
\label{spinor_review}

It was \'{E}lie Cartan who first has shown how the geometry of null vectors may be best dealt with in terms of spinors, nowadays defined as vectors of the representation space of Clifford algebras, whose components are equal (up to a sign) to the square root of linear combinations of null vector components (loosely speaking: spinors are square roots of null vectors)%
\footnote{He furthermore has shown how Euclidean geometry may be derived from spinor geometry which then could be the fundamental geometry of space-time and of natural phenomena also because it has the property of linearizing tensor equations, like, for example, Dirac equation linearizes Klein-Gordon equation.}%
. It is then appropriate to attempt to reformulate the MC problem in spinorial form.

Following Chevalley \cite{Chevalley 1954} spinors may be dealt with in the frame of Clifford algebras \cite{Benn 1987}, \cite{Budinich 1988}. Given a $2n$ dimensional complex space $\C^{2n}$, with Euclidean quadratic form and the corresponding Clifford algebra $Cl(2n)$, let $\gamma_{1} , \gamma_{2}, \ldots , \gamma_{2n}$ be the generators of $Cl(2n)$ with the property
\begin{displaymath}
\left[ \gamma_{j} , \gamma_{k} \right]_{+} := \gamma_{j} \gamma_{k}
+\gamma_{k} \gamma_{j} = 2 \delta_{jk} \Identity \qquad j,k=1,2, \ldots , 2n \; {\rm .}
\end{displaymath}
$Cl(2n)$ may be conceived as a direct sum of tensor spaces
\begin{displaymath}
Cl(2n)=V^{(0)} \oplus V^{(1)} \oplus \cdots \oplus V^{(2n)}
\end{displaymath}
and one can identify $V^{(1)}$ as the image of the vector space $\C^{2n}$ simply substituting the anticommutator with a scalar product and the $\gamma_{j}$ with the unit vectors ${\bf e}_{j}$ of a standard orthonormal basis of $\C^{2n}$ with coordinates $({\bf e}_{j})_{k} = \delta_{j k}$. 

A remark about notation: we indicated usual vectors in bold, so ${\bf v}$ represents a proper vector of $\C^{2n}$ while with $v$ we represent the corresponding element of the Clifford algebra $Cl(2n)$ belonging to tensor space $V^{(1)}$.

\smallskip

A spinor $\Phi$ is a vector belonging to the spaces $S$ of endomorphism of $Cl(2n) = \textrm{End} S$ and is defined by the Cartan's equation:
\begin{equation}
v \Phi = \left( \sum\limits_{j=1}^{2n} v_{j} \gamma_{j} \right) \Phi = 0
\label{formula_Cartan_equation}
\end{equation}
where $v_{j}$ are the orthonormal components of $v$ (and also of ${\bf v} \in \C^{2n}$) and $v \Phi$ is a Clifford product $v \Phi = v \JJ \Phi + v \wedge \Phi$.

\subsection{The Witt basis of $Cl(2n)$ and the Fock basis of the associated spinor $\Phi$}
\label{Witt_basis}
Let us define the null, or Witt, basis of $Cl(2n)$ as follows:
\begin{equation}
p_{j} =\frac{1}{2} \left( \gamma_{2j-1} +i\gamma_{2j} \right)
\quad \textrm{and} \quad
q_{j} =\frac{1}{2} \left( \gamma_{2j-1} -i\gamma_{2j} \right)
\quad j = 1,2, \ldots , n
\label{formula_Witt_basis}
\end{equation}
with the properties
\begin{equation}
\left[ p_{j} ,p_{k} \right]_{+} = \left[ q_{j} ,q_{k} \right]_{+} = 0
\quad \textrm{and} \quad
\left[ p_{j} ,q_{k} \right]_{+} = \delta_{jk} \Identity \; {\rm .}
\label{formula_Witt_basis_properties}
\end{equation}
With this basis $\C^{2n}$ is easily seen as the direct sum of two maximal TNP $P$ and $Q$ spanned by null vectors $\{{\bf p}_{j}\}$ and $\{{\bf q}_{j}\}$ respectively:
\begin{displaymath}
\C^{2n} = P \oplus Q \; {\rm ,}
\end{displaymath}
since $P \cap Q = \emptyset$ each vector ${\bf v} \in \C^{2n}$ may be expressed in the form ${\bf v} = \sum\limits_{i=1}^{n} \left( \alpha_{i} {\bf p}_{i} + \beta_{i} {\bf q}_{i} \right)$ with $\alpha_{i}$ and $\beta_{i}$ arbitrary complex numbers.

A spinor $\Phi \in S$, defined by Cartan equation (\ref{formula_Cartan_equation}), may be represented by Minimal Left Ideals (MLI) of $Cl(2n)$ \cite{Chevalley 1954}. Consider the $2^n$ MLI that form the Fock basis in spinor space \cite{Budinich 1989}
\begin{eqnarray}
& & \omega_{0} = {p}_{1} {p}_{2} \ldots {p}_{n}; \nonumber \\
& & \omega_{1} = {q}_{1} \omega_{0}, \quad \omega_{2} = {q}_{2} \omega_{0}, \quad \omega_{4} = {q}_{3} \omega_{0}, \quad \ldots , \quad \omega_{2^{n - 1}} = {q}_{n} \omega_{0}; \nonumber \\
& & \omega_{3} = {q}_{1} {q}_{2} \omega_{0}, \quad \omega_{5} = {q}_{1} {q}_{3} \omega_{0}, \quad \ldots ; \label{formula_Fock_basis} \\
& & \ldots\ldots \nonumber \\
& & \omega_{2^{n} - 1} = {q}_{1} {q}_{2} \ldots {q}_{n} \omega_{0} \nonumber
\end{eqnarray}
in which the indexes of the $q$'s always appear in ascending order and the interpretation of the $2^n$ values of the spinor index $s$ of $\omega_{s}$ is immediate thinking of $s$ as of a binary number of $n$ digits where the $j$-th digit from the right, taking the value $1$ or $0$, indicates wether $q_{j}$ is present or not in $\omega_{s}$. Any spinor $\Phi$ may be uniquely expressed in terms of the elements of the Fock basis (\ref{formula_Fock_basis})
\begin{equation}
\Phi = \sum\limits_{s=0}^{2^n - 1} \xi_{s} \omega_{s}
\label{formula_Fock_basis_sum}
\end{equation}
where the $\xi_{s}$ are the $2^n$ complex components of the spinor.

\subsection{Cartan equation in the Fock basis}
\label{Cartan_equation_1}
When we write the Cartan equation (\ref{formula_Cartan_equation}) in the basis, defined in (\ref{formula_Witt_basis}) and (\ref{formula_Fock_basis_sum}), we get
\begin{equation}
v \Phi = \left( \sum\limits_{i=1}^{n} \alpha_{i} p_{i} + \beta_{i} q_{i} \right) \left(\sum\limits_{s=0}^{2^n - 1} \xi_{s} \omega_{s}\right) = 0
\label{formula_Cartan_equation_FW_basis}
\end{equation}
and this equation can be read in two ways depending on wether $v$ or $\Phi$ plays the role of the unknown. For example, if $\Phi = \xi_0 \omega_{0}$, i.e. $\Phi = (\xi_0,0,\ldots,0)$%
\footnote{The spinor represented by the MLI $\omega_{0} = {p}_{1} {p}_{2} \ldots {p}_{n}$ was named standard by Cartan.}
it becomes
\begin{displaymath}
\left( \sum\limits_{i=1}^{n} \alpha_{i} p_{i} + \beta_{i} q_{i} \right) \xi_{0} \omega_{0} = \left( \sum\limits_{i=1}^{n} \alpha_{i} p_{i} + \beta_{i} q_{i} \right) \xi_{0} p_1 p_2 \ldots p_n = 0
\end{displaymath}
and, remembering that from (\ref{formula_Witt_basis_properties}) we have,
\begin{equation}
p_i p_i = 0 = q_i q_i \quad \quad p_i q_i = \Identity - q_i p_i
\label{formula_piqi_properties}
\end{equation}
one easily finds that the equation is satisfied, for $\xi_{0} \neq 0$, if, and only if, all the $\beta_i$ are zero. Moreover the equation holds for every value taken by the $\alpha_i$ i.e. for every point of the subspace $P = Span ({\bf p}_{1}, {\bf p}_{2}, \ldots, {\bf p}_{n})$. It is also simple to work this example the other way round i.e. given a subspace whose generic vector has the form ${\bf v} = \sum\limits_{i=1}^{n} x_{i} {\bf p}_{i}$ one finds that the spinor that satisfies (\ref{formula_Cartan_equation_FW_basis}) is of the form $\Phi = (\xi_0,0,\ldots,0) = \xi_0 \omega_{0}$ $\forall \xi_0 \in \C$.

This shows explicitly the correspondence, set up by Cartan equation (\ref{formula_Cartan_equation_FW_basis}), between spinors and TNP's, in this example between $\omega_{0}$ and the maximal TNP $P$ of $\C^{2n}$. Similarly one can find that the TNP corresponding to $\omega_{1}$ is $x_{1} q_{1} + \sum\limits_{i=2}^{n} x_{i} p_{i}$ and so on. More generally left-multiplying (\ref{formula_Cartan_equation_FW_basis}) by $v$ it becomes
\begin{displaymath}
v^2 \Phi = 0
\end{displaymath}
showing that $v^2$ is null for $\Phi \neq 0$ and equations (\ref{formula_Cartan_equation}) and (\ref{formula_Cartan_equation_FW_basis}) are linearizations of this relation. Let $v'$ be another solution of $v' \Phi = 0$ for the same $\Phi$, left-multiplying (\ref{formula_Cartan_equation_FW_basis}) by $v'$ we get
\begin{displaymath}
v' v \Phi = 0 \quad {\rm and \; similarly} \quad v v' \Phi = 0
\end{displaymath}
from which easily derives that
\begin{displaymath}
\left[ v', v \right]_{+} = 0
\end{displaymath}
so $v'$ and $v$, besides being null, are mutually orthogonal and thus form a TNP.

In general, given a spinor $\Phi$, all the vectors $v$ satisfying (\ref{formula_Cartan_equation_FW_basis}) are null and mutually orthogonal and define a TNP that we call $M_k(\Phi)$ where $k \le n$ indicates its dimensions. If $k = n$, that is, the dimension of $M_k(\Phi)$ is maximal, the corresponding spinor was called {\em simple} by Cartan \cite{Cartan 1937} and {\em pure} by Chevalley \cite{Chevalley 1954}, a name now prevailing in the literature. Each one of the $2^n$ spinors of the Fock basis (\ref{formula_Fock_basis}) is pure.

Consequently, from now on, when we indicate with $v$ the solution of $v \Phi = 0$ we actually refer to the entire subspace $M_k(\Phi)$ and not just to one of its vectors.

\bigskip

Pure spinors, as stressed by Cartan, are equivalent, up to a sign, to the corresponding maximal TNP whose null vectors may be bilinearly expressed in terms of them. This equivalence establishes a link between spinors and projective Euclidean geometry (of null vectors) which, being very simple and elegant, might have a crucial role for the explanation of several phenomena in physics. However there is a basic obstacle for setting in evidence this equivalence: while the dimensions of the TNP increase with $n$, that of the equivalent spinor increase with $2^{n}$ and consequently, for large $n$, their components will have to be subject to $O(2^{n})$ constraint relations. In order to overcome this difficulty, Cartan, when discussing the elegant properties and proving the theorems of pure spinor geometry, introduced the concept of standard pure spinors with only one component and therefore not subject to constraint relations.

\section{Spinorial formulation of the maximum clique problem}

We are now ready to give a spinorial formulation of the MC problem and start by introducing the vectors $\myconjugate{\bf z}_{i}$ of the matrix $\myconjugate{B}$ defined above, in the Witt basis of $Cl(2n)$ with the scalar product standing for an anticommutator:
\begin{displaymath}
\myconjugate{\bf z}_{i} = {\bf p}_{i} + \sum\limits_{j=1}^{n} \myconjugate{a}_{ij} {\bf q}_{j} \qquad i = 1,2, \ldots , n \; {\rm .}
\end{displaymath}
These $n$ vectors have the following properties (immediate to prove):
\begin{itemize}
\item belong to $\C^{2n}$ and are linearly independent (because of the ${\bf p}_{i}$);
\item are null, i.e. $\myconjugate{\bf z}_{i}^{\prime} \myconjugate{\bf z}_{i} = 0$ since $\myconjugate{a}_{ii} = 0$;
\item satisfy (\ref{formula_g}) for the complementary matrix, i.e. $\myconjugate{\bf z}_{i}^{\prime} \myconjugate{\bf z}_{j} = \myconjugate{a}_{ij}$.
\end{itemize}
In general they span an $n$ dimensional subspace $V$, which will be partially null. Precisely each  $\myconjugate{a}_{ij} = 0$ will imply the existence of a two dimensional TNP in $V$. $V$ will be totally null only if $\myconjugate{a}_{ij} = 0 \ \forall ij$ in which case $\myconjugate{\bf z}_{i} = {\bf p}_{i}$ and $V = P = Span ({\bf p}_{1}, {\bf p}_{2}, \ldots, {\bf p}_{n})$. Differently from (\ref{formula_g}) now $\myconjugate{B}$ is a $2n \times n$ matrix formed by $n$ linearly independent vectors such that $\myconjugate{A} = \myconjugate{B}^{\prime} \myconjugate{B}$.

\smallskip

To fully exploit the spinorial formulation we will consider the $\myconjugate{\bf z}_{i}$ vectors as representative of the subspace they induce i.e. $Span ({\bf p}_{i}, \myconjugate{a}_{i1} {\bf q}_{1}, \ldots,  \myconjugate{a}_{in} {\bf q}_{n})$ of dimension $\sum\limits_{j=1}^{n} \myconjugate{a}_{ij} + 1$. We do this introducing in the definition arbitrary coefficients $\alpha$
\begin{equation}
\myconjugate{\bf z}_{i} = \alpha_i {\bf p}_{i} + \sum\limits_{j=1}^{n} \myconjugate{a}_{ij} \alpha_j {\bf q}_{j} \qquad i = 1,2, \ldots , n
\label{formula_zi_def}
\end{equation}
and we can always get back the representative vectors setting all $\alpha = 1$.

\smallskip

The equation $\myconjugate{\bf Z}^{2} = 0$, representing the constraints of the MC problem in (\ref{formula_j}), may be linearized formulating the problem in spinorial form:
\begin{equation}
\myconjugate{\bf Z} \Phi = \myconjugate{B} {\bf x} \Phi = \left( \sum\limits_{i=1}^{n} x_{i} \myconjugate{\bf z}_{i} \right) \Phi = 0
\label{formula_t}
\end{equation}
or with $\myconjugate{\bf z}_{i} $ from (\ref{formula_zi_def})
\begin{equation}
\left[ \sum\limits_{i=1}^{n} x_{i} \left( \alpha_i {\bf p}_{i} + \sum\limits_{j=1}^{n} \alpha_j \myconjugate{a}_{ij} {\bf q}_{j} \right) \right] \Phi = 0
\label{formula_u}
\end{equation}
of the form (\ref{formula_Cartan_equation_FW_basis}). In this equation, in general, $x_{i}$ must be interpreted as complex variables, restricted to values in $\{0, 1\}$ in the traditional formulation of the MC problem (\ref{formula_j}).

We thus have a set of $n$ vectors $\myconjugate{\bf z}_{i}$ defining an $n$-dimensional subspace of $\C^{2 n}$ and we will look for the spinors $\Phi$ that satisfy (\ref{formula_t}).

\subsection{Some properties of Cartan equation}
\label{Cartan_equation_2}
Before analyzing in detail (\ref{formula_u}), devoted to graphs, we step back to the general form of Cartan equation (\ref{formula_Cartan_equation_FW_basis}) and derive some of its properties. First we study the case in which the TNP is not maximal.

\begin{MS_Proposition}
\label{prop_dim}
Given a TNP of dimension $k \le n$, the corresponding spinor $\Phi$, solution of the Cartan equation (\ref{formula_Cartan_equation_FW_basis}), has at least $2^{n - k}$ non-zero coordinates in the Fock basis.
\end{MS_Proposition}

Without loss of generality we take $Span ({\bf p}_{1}, {\bf p}_{2}, \ldots, {\bf p}_{k})$ as a TNP of dimension $k$, since given any TNP of dimension $k$ it is always possible, by a proper choice of the basis, make it coincide with $Span ({\bf p}_{1}, {\bf p}_{2}, \ldots, {\bf p}_{k})$ (see e.g. \cite{Budinich 1989}).

We will prove the proposition by induction: we already know that when the TNP is maximal, i.e. of dimension $n$ as in the example in paragraph~\ref{Cartan_equation_1}, the corresponding spinor can have one component%
\footnote{Actually it may have at most two components, see proposition~$5$ in \cite{Budinich 1989}.}
and thus satisfies the proposition.

Let us suppose now that we have a TNP of dimension $k$ and that the corresponding spinor $\Phi$ has $m$ nonzero components: we will show that when reducing the dimension of the TNP to $k-1$ the new spinor has {\em at least} $2 m$ components. So let us suppose that our spinor $\Phi$ has $m$ nonzero components and satisfy
\begin{displaymath}
\left( \sum\limits_{i=1}^{k} x_{i} {p}_{i} \right) \Phi = \left( \sum\limits_{i=1}^{k - 1} x_{i} {p}_{i} + x_{k} {p}_{k} \right) \Phi = 0 \; {\rm .}
\end{displaymath}
Since this relation holds for any value of the $x_i$ it must hold in particular for $x_k = 0$ so that we have
\begin{displaymath}
\left( \sum\limits_{i=1}^{k - 1} x_{i} {p}_{i} \right) \Phi = 0
\end{displaymath}
from which necessarily derives ${p}_{k} \Phi = 0$. This implies that ${q}_{k}$ does not appear in all the $m$ $\omega$'s that are the nonzero components of $\Phi$. Assuming the contrary we could write
\begin{displaymath}
x_{k} {p}_{k} \sum\limits_{s} \xi_{s} \omega_{s} = x_{k} {p}_{k} \left( \sum\limits_{s \in \{ q_k \}} \xi_{s} \omega_{s} + \sum\limits_{s \in \{ \bar{q}_k \}} \xi_{s} \omega_{s} \right) = x_{k} {p}_{k} \sum\limits_{s \in \{ q_k \}} \xi_{s} \omega_{s} = 0
\end{displaymath}
where by $s \in \{ q_k \}$ we indicate the subset of the $m$ values of $s$ such that the term $q_k$ do appear in $\omega_{s}$ and by $s \in \{ \bar{q}_k \}$ the complementary subset in which the term $q_k$ do not appear; obviously this second sum vanish when left multiplied by $p_k$. The generic term of the surviving sum can be easily calculated with (\ref{formula_piqi_properties}) and reduces to
\begin{displaymath}
x_{k} \sum\limits_{s \in \{ q_k \}} \xi_{s} (-1)^{l_s} (\Identity - q_k p_k) \omega_{s - 2^k} = x_{k} \sum\limits_{s \in \{ q_k \}} \xi_{s} (-1)^{l_s} \omega_{s - 2^k} = 0
\end{displaymath}
when ${q}_{k}$ is the $l_s$-th of the $q$'s present in $\omega_{s}$. But this relation cannot hold because it would imply that the components of the Fock basis are linearly dependent (remark that $\omega_{s - 2^k}$ are all different). Thus we proved that none of the $m$ components of $\Phi$ can contain ${q}_{k}$.

Returning to our argument we observe that when we reduce the size of the TNP by $1$, setting $x_k \equiv 0$, we have
\begin{displaymath}
\left( \sum\limits_{i=1}^{k - 1} x_{i} {p}_{i} \right) \Phi = 0
\end{displaymath}
but also
\begin{displaymath}
\left( \sum\limits_{i=1}^{k - 1} x_{i} {p}_{i} \right) (\Phi + q_{k} \Phi) = 0
\end{displaymath}
and since no component of $\Phi$ contains $q_{k}$ all components of $q_{k} \Phi$ are different from those of $\Phi$ and the spinor $\Phi + q_{k} \Phi$ has $2 m$ nonzero components. This concludes the induction argument proving the proposition. $\Box$

An immediate consequence of the arguments of the proof is that $\Phi$ relative to $Span ({\bf p}_{1}, {\bf p}_{2}, \ldots, {\bf p}_{k})$ cannot have components with any of the $\{ {q}_{1}, {q}_{2}, \ldots, {q}_{k} \}$. This allows us to write $\Phi$ explicitly: its components are all and only the $2^{n-k}$ not containing any element of $\{ {q}_{1}, {q}_{2}, \ldots, {q}_{k} \}$. This generalizes easily to the case of a TNP of size $k$ of the more general form
\begin{displaymath}
\left( \sum\limits_{i} x_{i} {p}_{i} + \sum\limits_{j} x_{j} {q}_{j} \right) \Phi = 0
\end{displaymath}
the corresponding spinor components are all and only the $2^{n-k}$ components not containing any element of $\{ {q}_{i} \}$ and containing all of the elements of $\{ {q}_{j} \}$.

\bigskip

We now consider the more general case of a TNP formed by the span of $k$ null vectors $v$ and show a way to express the corresponding spinor.

\begin{MS_Proposition}
\label{prop_spinor_build}
Given a TNP $V = Span (v_1,\ldots,v_k)$, of dimension $k$, the corresponding spinor $\Phi (v_1,\ldots,v_k)$, satisfying the Cartan equation (\ref{formula_Cartan_equation_FW_basis}), can be calculated with
\begin{equation}
\Phi (v_1,\ldots,v_k) = v_1 \ldots v_k \Phi (\Identity)
\label{formula_gen_spinor_def}
\end{equation}
where with $\Phi (\Identity)$ we represent the most general spinor (\ref{formula_Fock_basis_sum}) expressed in the Fock basis.
\end{MS_Proposition}

We start by proving the simpler case in which there is just one vector and moreover this vector coincides with one of the basis, i.e. $v = p_j$ and we prove that
\begin{displaymath}
\Phi (p_j) = p_j \Phi (\Identity)
\end{displaymath}
To calculate $p_j \Phi (\Identity)$, as in the previous proof, we split the sum over the Fock basis into two parts and get
\begin{eqnarray}
p_j \Phi (\Identity) & = & p_j \sum\limits_{s \in \{ q_j \}} \xi_{s} \omega_{s} + p_j \sum\limits_{s \in \{ \bar{q}_j \}} \xi_{s} \omega_{s} = p_j \sum\limits_{s \in \{ q_j \}} \xi_{s} \omega_{s} = \nonumber \\
& = & \sum\limits_{s \in \{ q_j \}} \xi_{s} (-1)^{l_s} (\Identity - q_j p_j) \omega_{s - 2^j} = \sum\limits_{s \in \{ q_j \}} \xi_{s} (-1)^{l_s} \omega_{s - 2^j} = \nonumber \\
& = & \sum\limits_{s \in \{ \bar{q}_j \}} \xi_{s + 2^j} (-1)^{l_{s + 2^j}} \omega_{s} \nonumber
\end{eqnarray}
that show that $p_j \Phi (\Identity)$  obviously satisfy $p_j \Phi (p_j) = 0$ and has for components the $2^{n - 1}$ in which $q_j$ do not appear in $\omega_{s}$ and thus, given the arbitrary values of the coefficients $\xi_s$, represents the most general expression for $\Phi (p_j)$, as shown in proposition~\ref{prop_dim}.

To prove the general proposition we proceed by induction and first extend the proof to a more general null vector $v_1 = \sum\limits_{i} \alpha_{i} {p}_{i} + \sum\limits_{j} \beta_{j} {q}_{j}$
\begin{displaymath}
\Phi (v_1) = v_1 \Phi (\Identity) = \sum\limits_{i} \alpha_{i} \Phi ({p}_{i}) + \sum\limits_{j} \beta_{j} \Phi ({q}_{j})
\end{displaymath}
and
\begin{displaymath}
v_1 \Phi (v_1) = v_1^2 \Phi (\Identity) = 0
\end{displaymath}
since, by hypothesis, $v_1$ is null. Now suppose that we already have
\begin{displaymath}
\Phi (v_1,\ldots,v_{j-1}) = v_1 \ldots v_{j-1} \Phi (\Identity)
\end{displaymath}
and we add $v_{j}$ that form a TNP with previous vectors, we have
\begin{displaymath}
\left( x_{j} v_{j} + \sum\limits_{i=1}^{j-1} x_{i} {v}_{i} \right) v_j \Phi (v_1,\ldots,v_{j-1}) = - v_{j} \left( \sum\limits_{i=1}^{j-1} x_{i} {v}_{i} \right) \Phi (v_1,\ldots,v_{j-1}) = 0
\end{displaymath}
where we have used the relations $v_{j}^2 = 0$ and $v_i v_j = - v_j v_i$ for $i < j$ deriving from the hypothesis that the $v$'s form a TNP. This completes the proof showing also that the order of multiplication of $v$'s in (\ref{formula_gen_spinor_def}) is irrelevant since it can only affect the global sign of the spinor. $\Box$

\bigskip

We conclude with the general case:

\begin{MS_Proposition}
\label{prop_non_null_vectors}
Given $V := Span ({v}_1,\ldots,{v}_k)$, there exists a spinor $\Phi$, satisfying the Cartan equation for all values of the coefficients ${x}_1,\ldots,{x}_k$
\begin{displaymath}
\left( \sum\limits_{i=1}^k x_{i} {v}_{i} \right) \Phi = 0
\end{displaymath}
if, and only if, $V$ is a TNP.
\end{MS_Proposition}

To prove it let us suppose the contrary, i.e. that there exists a non null vector $v$ and $\Phi \neq 0$ satisfying Cartan equation; left multiplying by $v$ we would get
\begin{displaymath}
v^2 \Phi = 0
\end{displaymath}
but since $v^2 \neq 0$ this implies $\Phi = 0$ contradicting the initial hypothesis. On the other hand, for any null vector $v$ the spinor $\Phi (v) = v \Phi (\Identity)$ satisfies the Cartan equation since $v \Phi (v) = v^2 \Phi (\Identity) = 0$. $\Box$

\subsection{Back to the graph maximum clique problem}

We get back to our form of Cartan equation (\ref{formula_u}) with an example: let us suppose that $\myconjugate{a}_{1 2} = 0$, this means that $\myconjugate{\bf z}_{1}$ and $\myconjugate{\bf z}_{2}$ form a TNP and, setting $x_3 = x_4 = \ldots = x_n = 0$, with (\ref{formula_zi_def}) and proposition~\ref{prop_spinor_build} we get
\begin{displaymath}
\left( x_1 \myconjugate{\bf z}_{1} + x_2 \myconjugate{\bf z}_{2} \right) \Phi(\myconjugate{\bf z}_{1} \myconjugate{\bf z}_{2}) = \left( x_1 \myconjugate{\bf z}_{1} + x_2 \myconjugate{\bf z}_{2} \right) {\bf p}_{1} {\bf p}_{2} \ldots {\bf q}_{i} \ldots \Phi(\Identity) = 0 \; {\rm .}
\end{displaymath}
where with the notation $\ldots {\bf q}_{i} \ldots$ we indicate all different ${\bf q}_{i}$ that appear in $\myconjugate{\bf z}_{1}$ and $\myconjugate{\bf z}_{2}$.
This example shows that it is simple to get particular solutions to (\ref{formula_u}) the real problem being to find the {\em set of all solutions}, for which we have the following
\begin{MS_Proposition}
\label{isomorphism_cliques_spinors}
The set of nonzero spinors that solve the Cartan equation (\ref{formula_u}) is isomorphic to the set of cliques of $A$.
\end{MS_Proposition}
Given a solution of (\ref{formula_u}) with $\Phi \ne 0$ we will have corresponding values for the coefficients $x_1, \ldots, x_n$. For all $x_i \ne 0$ we can redefine the arbitrary $\alpha$ coefficients in (\ref{formula_u}) so that $x_i = 1$ and the solution can be written in the form $x_i \in \{0,1\}$. The $\myconjugate{\bf z}_{i}$ corresponding to $x_i = 1$ form necessarily a clique since for any couple of them their scalar product is null. Viceversa given a clique $\myconjugate{\bf z}_{i_1}, \ldots, \myconjugate{\bf z}_{i_k}$ the corresponding coefficients $x_i \in \{0,1\}$ and the corresponding spinor $\Phi(\myconjugate{\bf z}_{i_1}, \ldots, \myconjugate{\bf z}_{i_k})$ satisfy the Cartan equation. $\Box$

We can now reformulate our initial MC problem (\ref{formula_j}): it will correspond to that solution of (\ref{formula_u}) with the maximum intersection with $P$, i.e.
\begin{equation}
k = \max_{\Phi : \left( \sum\limits_{i=1}^{n} x_{i} \myconjugate{\bf z}_{i} \right) \Phi = 0}
{\rm dim} \left( P \cap M_{} (\Phi) \right) \; {\rm .}
\label{formula_mc_spinorial}
\end{equation}
This shows also that the problem of finding all possible spinor solutions of (\ref{formula_u}) is NP-complete since, given the set of all solutions, one gets also the solution of the MC problem. With respect to the MC formulation (\ref{formula_j}) we remark two main differences: the first is that the demanding restriction ${\bf x} \in \{0,1\}^n$ can be relaxed since all solutions of (\ref{formula_u}) necessarily have binary $x_i$. The second is that the quadratic constraint $\myconjugate{\bf Z}^{2} = 0$ of (\ref{formula_j}) is linearized here to $\myconjugate{\bf Z} \Phi = 0$.

\subsection{Definition of the spinor $\Psi(\myconjugate{A})$ corresponding to graph $A$}

Even if we cannot determine in general the set of all solutions to our problem (that would mean solving an NP-complete problem) we can try to better characterize the set of spinors satisfying (\ref{formula_u}). Let us define a null vector $\myconjugate{\bf Z}_k$, sum of $k$ vectors $\myconjugate{\bf z}_{i}$,
\begin{displaymath}
\myconjugate{\bf Z}_{k} = \myconjugate{\bf z}_{j_1} + \myconjugate{\bf z}_{j_2} + \cdots + \myconjugate{\bf z}_{j_k} = Span ({\bf p}_{j_1}, {\bf p}_{j_2}, \ldots, {\bf p}_{j_k}, \ldots {\bf q}_{i} \ldots)
\end{displaymath}
as {\em saturated} if it is null and if  no other $\myconjugate{\bf z}_{i}$ can be added to it without destroying its nullness. In other words $\myconjugate{\bf Z}_k$ is saturated if, in its expression, appear exactly $n$ of the ${\bf p}_{i}$ and ${\bf q}_{i}$ vectors all with different indexes. A clique is said to be {\em maximal} if no other vertex can be added obtaining a new clique (obviously a maximum clique is also maximal). We show that the concepts are identical:

\begin{MS_Proposition}
The set of saturated vectors formed with the $\myconjugate{\bf z}_{i}$ is isomorphic to that of the maximal cliques of the corresponding graph.
\end{MS_Proposition}

Let us suppose that the null vector $\myconjugate{\bf Z}_k$ is saturated, obviously the set of its ${\bf p}_{j_{i}}$ vectors uniquely identify a subgraph. From its nullness we have $\myconjugate{\bf z}_{j_{i}}^{\prime} \myconjugate{\bf z}_{j_{l}} = \myconjugate{a}_{j_{i}j_{l}} = 0$ for every ${j_{i}j_{l}}$ and the identified subgraph is a clique. To prove that this clique is also maximal let us suppose the contrary, i.e. that the vertex $j_{k+1}$ can be added to it obtaining a larger clique. Then we would necessarily have $\myconjugate{a}_{j_{i}j_{k+1}} = 0$ for all $j_{i}$ that would mean that the vector ${\bf q}_{j_{k+1}}$ would be missing from $\myconjugate{\bf Z}_k$ violating the hypothesis that $\myconjugate{\bf Z}_k$ is saturated.

To prove the second part of the proposition let us suppose that we have a maximal clique identified by a vector ${\bf x}_k \in \{0,1\}^n$; the vector $\myconjugate{\bf Z}_k = \myconjugate{B} {\bf x}_k$ is null since $0 = \myconjugate{a}_{j_{i}j_{l}} = \myconjugate{\bf z}_{j_{i}}^{\prime} \myconjugate{\bf z}{j_{l}}$ for every ${j_{i}j_{l}}$. Also $n - k$ of the ${\bf q}_{i}$ appear in it otherwise, as in the preceding part, one can easily contradict the hypothesis that the starting clique is maximal. It follows that $\myconjugate{\bf Z}_k$ is saturated. $\Box$

Each saturated vector $\myconjugate{\bf Z}_k$, made up by $k$ $\myconjugate{\bf z}_i$'s, can be thought as a $k$-dimensional TNP (remember $\myconjugate{\bf z}_i$'s are linearly independent) but also, using the more general definition (\ref{formula_zi_def}), one can use the ${\bf p}$'s and ${\bf q}$'s that appear in $\myconjugate{\bf Z}_k$ to build a maximal TNP whose spinor is pure and given by
\begin{equation}
\Phi(\myconjugate{\bf Z}_k) = {\bf p}_{j_1} \ldots {\bf p}_{j_k} \ldots {\bf q} \ldots \Phi(\Identity) := \omega_{\myconjugate{Z}_k}
\label{formula_spinor_saturated_vector_components}
\end{equation}
so that each saturated vector uniquely identifies one of the pure spinors of the Fock basis (\ref{formula_Fock_basis}). Thus, in our formulation, every maximal clique corresponds to a saturated vector $\myconjugate{\bf Z}_k$ which in turn identifies one of the components of the Fock basis (\ref{formula_Fock_basis}).

We remark that the $M(\Phi(\myconjugate{Z}_k))$ has dimension $n$ and the set of the ${\bf p}_{i}$ of $\myconjugate{\bf Z}_k$ always allow to indicate unambiguously the maximal clique associated with it. In other words, each maximal TNP $M(\Phi(\myconjugate{Z}_k))$ contain one and only one of the saturated vectors and thus just one maximal clique.

\smallskip

We show now the equivalence of the formulation of the Cartan equation

\begin{MS_Proposition}
Given a set of $x_{j_i}$ that give a solution of the Cartan equation (\ref{formula_u}) and calling $\myconjugate{\bf Z}_1,\ldots,\myconjugate{\bf Z}_p$ all the saturated vectors such that each of them contains all the $\myconjugate{\bf z}_{j_i}$ we are considering, then
\begin{displaymath}
\left( \sum\limits_{i=1}^k x_{j_i} \myconjugate{\bf z}_{j_i} \right) \left( \sum\limits_{l=1}^p \Phi(\myconjugate{\bf Z}_{l}) \right) = 0
\end{displaymath}
\end{MS_Proposition}

First of all we observe that $p \ge 1$ since the set of $\myconjugate{\bf z}_{j_i}$ form a TNP and are thus contained in at least one maximal clique how shown in the constructive proof of proposition~\ref{maximal_cliques_identify_graph}. By (\ref{formula_spinor_saturated_vector_components}) all $\Phi(\myconjugate{\bf Z}_{l})$ contain all the $\myconjugate{\bf z}_{j_i}$ of the first sum and the proposition is proved. $\Box$

Since, as we prove in the Appendix, each graph is uniquely identified by the set of its maximal cliques it is possible to define uniquely a spinor $\Psi(\myconjugate{A})$ associated to a given graph $A$: the set of all its maximal cliques defines uniquely a set of saturated vectors $\{ \myconjugate{\bf Z}_{l} \}$. This set defines in turn a corresponding set $\{ \omega_{{\bf Z}_{l}} \}$ in the Fock basis and therefore also a spinor $\Psi(\myconjugate{A})$ of the form (\ref{formula_Fock_basis_sum}) uniquely defined by $\myconjugate{A}$:
\begin{equation}
\label{spinor_graph_definition}
\Psi(\myconjugate{A}) = \sum\limits_{l} \xi_{l} \Phi(\myconjugate{\bf Z}_{l}) = \sum\limits_{l} \xi_{l} \omega_{\myconjugate{\bf Z}_{l}} \; {\rm .}
\end{equation}

We note that in this formulations not all components are different from zero for all the values of the $x_i$ and we can render explicit this characteristic adding to each components a product of Kronecker delta that set it to zero when one of the components having positive scalar products with one of the $\myconjugate{\bf Z}_{l}$ is present so, calling $Q_l$ the set of the indices of $q_i$ appearing in $\myconjugate{\bf Z}_{l}$, we get:
\begin{displaymath}
\Psi(\myconjugate{A}) = \sum\limits_{l} \xi_{l} \omega_{\myconjugate{\bf Z}_{l}} \Delta (l, x_1, x_2, \ldots, x_n)
\end{displaymath}
with
\begin{displaymath}
\Delta (l, x_1, x_2, \ldots, x_n) = \prod_{j \in Q_l} \delta_{x_j,0}
\end{displaymath}
and with this definition we can now rewrite (\ref{formula_t}) as
\begin{displaymath}
\left( \sum\limits_{i=1}^{n} x_{i} \myconjugate{\bf z}_{i} \right) \Psi(\myconjugate{A}) = 0 \; {\rm .}
\end{displaymath}

For example if $\myconjugate{A} = 0$, $\Psi(\myconjugate{A}) = \omega_{0}$ while for $\myconjugate{A} = J - \Identity$,  $\Psi(\myconjugate{A}) = \xi_{1} \omega_{1} + \xi_{2} \omega_{2} + \xi_{4} \omega_{4} + \cdots + \xi_{2^{n-1}} \omega_{2^{n-1}}$. In general $\Psi(\myconjugate{A})$ will have a number of non-zero components lower than $2^{n}$ but there exist graphs with an exponential number of maximal cliques.

While $\myconjugate{A}$ uniquely determines $\Psi(\myconjugate{A})$, not every spinor may be conceived as generated by a graph. As an example a spinor $\Psi(\myconjugate{A})$ may not obviously contain components in $\{ \omega_{1}, \omega_{3}, \omega_{7}, \ldots \}$. The spinors $\Psi(\myconjugate{A})$ generated by graphs build up a subclass of spinors whose properties should be further analyzed and they will be certainly interesting also for other fields of application of pure spinor geometry, one can conjecture that they fall in the class of generalized spinors studied in \cite{Trautman 1994}.

\section{Conclusions}
Spinors were discovered by Cartan in $1913$ and soon after introduced in physics for the representation of the electron (and fermions) by Dirac and Weyl. However the geometry of pure spinors was practically ignored after the publication of Chevalley book in $1954$ \cite{Chevalley 1954}. The main motivation is that general spinors are too difficult to deal with because of the exponentially many constraint relations, for large $n$.

However now, after $50$ years, the scenario may change and pure spinors might attract the attention of both theoretical physicists and mathematicians. The main reason of this change is that theoretical physics, since several decades, is facing insurmountable difficulties in some of its central sectors like the quantization of the gravitational field or the explanation of some aspects of elementary particles phenomenology (origin of charges, families, etc.). Recently, pure spinors have been discovered to allow to overcome, somehow miraculously, some of these difficulties \cite{Berkovits} and to allow to shed some light on some aspects of the obscure phenomenology of elementary particles \cite{Budinich 2002}.

As shown by Cartan the geometry of pure spinors is correlated to that of null vectors and totally null planes, and shares the elegance and simplicity of projective geometry. We have shown that null vectors and TNP are deeply connected to graphs and formulating the MC problem in spinorial language establishes a bridge between these two worlds that, we hope, will allow to cross-fertilize both the fields. If and when, the certainly rich and elegant geometry of the pure spinors will be better known, it might contribute also to the MC problem, once this is formulated in the frame of that geometry, as proposed here.

In this paper, establishing a correspondence between totally null planes and maximal cliques, we have been able to reformulate neatly (\ref{formula_mc_spinorial}) the maximum clique problem and to define spinors corresponding to graphs (\ref{spinor_graph_definition}). Another interesting aspect, which emerges already in this preliminary approach, is that the pure spinor defined by graphs might belong to a subclass less prone to constraint relations.

\appendix
\section{Appendix}

Any symmetric matrix like $A$ may be expressed as $A = B^2$ where $B$ is a complex symmetric matrix. Since $A$ is symmetric its eigenvalues are real and it can be diagonalized $A = O^{\prime} \Lambda O$ with $O^{\prime} O = O O^{\prime} = \Identity$ where $\Lambda$ is the diagonal matrix of the eigenvalues $\lambda_i$ and $O^{\prime}$ is the orthogonal, real, matrix of the eigenvectors. Then, as it is easy to verify, a possible definition of the ``square root'' of $A$ is $B=O^{\prime} \sqrt{\Lambda } O$ where $\sqrt{\Lambda}$ is the diagonal matrix whose elements are the square roots of the eigenvalues of $A$ \cite{Horn 1992}.

We observe that, unless $A$ is semipositive definite ($\lambda_i \ge 0$), $B$ is complex and the choice of the signs of the diagonal elements of $\sqrt{\Lambda}$ is arbitrary so, in general, there are at least $2^n$ different possible $B$ satisfying (\ref{formula_f}). Moreover when $A$ has multiple eigenvalues, there are infinitely many possible choices of the corresponding eigenvectors, and there are, accordingly, infinitely many possible choices for $B$.

We conclude by proving the following
\begin{MS_Proposition}
\label{maximal_cliques_identify_graph}
Each graph is uniquely identified by the set of its maximal cliques.
\end{MS_Proposition}

To prove this assertion we provide a constructive algorithm to build $A$ from the set of its maximal cliques. One starts from $A = 0$ and add to it (in a Boolean logic fashion) all the links of each maximal clique; it is sufficient to add only the links between the nodes of a maximal clique and these links are all known, since one knows the subset of vertices that forms a maximal clique. This procedure brings to the adjacency matrix of the graph since each link appears in at least one of the maximal cliques. This last statement is proved observing that each maximal clique can be built starting from any link, and the corresponding two nodes, and adding to them, one at the time, other fully connected nodes. This proves that any link must appear in at least one maximal clique and thus the proposition. $\Box$

\end{document}